\documentclass[a4paper]{article}
\usepackage[ruled,vlined]{algorithm2e}
\usepackage{INTERSPEECH2019}
\usepackage{amsmath}

\usepackage{booktabs}
\usepackage{multirow}
\usepackage{colortbl}
\usepackage{diagbox}
\usepackage{pgfplots}
\usepackage{float}
\usepackage{makecell}
\newcolumntype{?}{!{\vrule width 1pt}}

\pgfplotsset{compat=1.14}

\usepackage{array}
\newcolumntype{L}[1]{>{\raggedright\let\newline\\\arraybackslash\hspace{0pt}}m{#1}}
\newcolumntype{C}[1]{>{\centering\let\newline\\\arraybackslash\hspace{0pt}}m{#1}}
\newcolumntype{R}[1]{>{\raggedleft\let\newline\\\arraybackslash\hspace{0pt}}m{#1}}

\title{Improved Speech Separation with Time-and-Frequency Cross-domain Joint Embedding and Clustering}
\name{Gene-Ping Yang$^1$, Chao-I Tuan$^2$, Hung-Yi Lee$^3$, Lin-shan Lee$^4$}
\address{
  Graduate Institute of Networking and Multimedia, National Taiwan University}
\email{r06944010@ntu.edu.tw$^1$, chaoi111.t@gmail.com$^2$, tlkagkb93901106@gmail.com$^3$, lslee@gate.sinica.edu.tw$^4$}

\begin{document}

\makeatletter
\newcommand*{\rom}[1]{\expandafter\@slowromancap\romannumeral #1@}
\makeatother

\maketitle
\begin{abstract}

% We proposed a hybrid framework that fuses temporal and spectral features in audio representations to improve the performance of speaker-independent source separation. We integrate both domain features by combining their encoded representations, finding our model capable of utilizing information in a multi-domain fashion of the original input mixtures and obtain significant progress on the performance of the challenging Wall Street Journal dataset (WSJ0) with two speaker mixtures. By further adapting classic clustering approach on the multi-domain features enforce the partition of the features close to the desired speaker assignments. Experimental results show that our model yields better performance and robustness, and achieve state-of-the-art performance compared to previous methods with flying colours.

Speech separation has been very successful with deep learning techniques. Substantial effort has been reported based on approaches over spectrogram, which is well known as the standard time-and-frequency cross-domain representation for speech signals. It is highly correlated to the phonetic structure of speech, or "how the speech sounds" when perceived by human, but primarily frequency domain features carrying temporal behaviour. Very impressive work achieving speech separation over time domain was reported recently, probably because waveforms in time domain may describe the different realizations of speech in a more precise way than spectrogram. In this paper, we propose a framework properly integrating the above two directions, hoping to achieve both purposes. We construct a time-and-frequency feature map by concatenating the 1-dim convolution encoded feature map (for time domain) and the spectrogram (for frequency domain), which was then processed by an embedding network and clustering approaches very similar to those used in time and frequency domain prior works. In this way, the information in the time and frequency domains, as well as the interactions between them, can be jointly considered during embedding and clustering. Very encouraging results (state-of-the-art to our knowledge) were obtained with WSJ0-2mix dataset in preliminary experiments.
 
\end{abstract}

\section{Introduction}

Human beings are able to focus on the voice produced by a single speaker when conversing with a particular individual in a crowded and noisy environment. %with lots of speakers speaking simultaneously.
This requires the ability to extract the desired voice from some mixed audio signal. This so-called cocktail party problem has been shown to be difficult for computers. Very impressive results have been achieved when the speakers are known in advance, but the task remains challenging when the speakers in the mixed voice are not known, referred as the speaker-independent source separation problem. Substantial effort has been made on this problem, obviously because good solutions to it may lead to good contributions to many downstream tasks such as speech recognition \cite{kim2017joint}, speaker identification \cite{nagrani2017voxceleb}, and audio classification \cite{gemmeke2017audio} in noisy environment.

Deep learning techniques have accomplished a big step forward on this speech separation task. The Deep Clustering (DPCL) technique \cite{hershey2016deep} was a good such example. It successfully coped with this problem to a good extent by projecting each element in the mixture magnitude spectrogram to a high-dimensional embedding space which is more discriminative for speaker partitioning. Various related deep learning approaches have been proposed and showed great success in enhancement \cite{pascual2017segan, rethage2018wavenet, choi2019phase} and separation \cite{wang2014training, chen2017deep, yu2017permutation, luo2017deep, luo2018tasnet1} tasks, although many techniques reported at that time consisted of multiple stages separately optimized under different criteria, such as signal representation and embeddings, embedding clustering for speaker assignments \cite{hershey2016deep, isik2016single}, mask generation over mixture magnitude spectrogram \cite{wang2014training, luo2018speaker}, and phase approximation approaches \cite{wang2018alternative, wang2018end, wang2018deep}. End-to-end approaches then became popular later on, in which all different stages with different functions were jointly trained \cite{miao2015eesen, chan2015listen}.

For most methods performed over the magnitude spectrogram, the ignorance of the phase of the individual sources inevitably distorted the time domain signals. Moreover, predicting masks for each source also caused mismatch for the individual signals. These problems remained even with great effort made. A very impressive phase estimation approach was proposed recently \cite{wang2018deep} based on a trigonometric perspective over the magnitude spectrogram. This approach achieved great progress over previous methods, offering a signal-to-distortion-ratio improvement SDR$_i = 15.6$ on the publicly available datasets WSJ0-2mix \cite{hershey2016deep}, which seems to be the recent state-of-the-art on the task.

On the other hand, different from the above mentioned methods operated over the spectrograms, a surprisingly successful approach was reported recently called TasNet \cite{luo2018tasnet}, which directly handled the problem over the time domain signals and achieved superior performance with SDR$_i = 15.0$ dB on WSJ0-2mix dataset. It contained an encoder module, a separation module and a decoder module, where the separation module consisted of multiple blocks of dilated convolutional layers similar to Wavenet \cite{oord2016wavenet}, but with fewer parameters and less computation due to the adoption of depthwise separable convolution previously proposed \cite{howard2017mobilenets}. 

To the best of our knowledge, existing approaches for the considered problem have taken either time or frequency domain representations as input, both achieving very good and very close performance. Obviously, both representations possess their respective advantages: the robustness of spectrograms in frequency domain, and the sophisticated but fine structures of time domain signals. The spectrogram is highly correlated to the phonetic structure of speech, or  `` how the speech sounds'' when perceived by human. But the waveform in time domain describes the various realizations of the sound in a more precise way.

In this paper, we try to integrate both time and frequency domain features together, with the hope to take the advantages of both domains. We construct a time-and-frequency feature map by concatenating features for both time and frequency domains, and perform cross-domain joint embedding and clustering over this feature map, so the model can learn signal behavior in both domains as well as the cross-domain correlations. We make part of the approach similar to TasNet \cite{luo2018tasnet}, which has fewer parameters with large receptive field due to the dilated convolutional layers. We also adopt the previously proposed clustering method for mask estimation \cite{luo2018speaker}, which directly predicts masks for each source in the mixture from feature embeddings. Such a model structure is also in good parallel to the insight offered by a recently reported work \cite{ravanelli2018speaker}. As will be shown below, very encouraging performance (state-of-the-art to our knowledge) was obtained on WSJ0-2mix dataset in preliminary experiments.

\section{Proposed Approach}

% \begin{figure*}[t]
%   \centering
%   \includegraphics[width=0.8\linewidth]{IS2019_paper_kit/LaTeX/diagram3.png}
%   \caption[HybridNet block diagram.]
%   {HybridNet block diagram. An encoder extract features from input signals, and the embedding network projects each extracted features to a high-dimensional vector, following by clustering which calculates masks for each source speaker. A decoder reconstruct the time-domain signals of each source from the multiplication of masks and the encoded features.} 
%   \label{fig:diagram}
% \end{figure*}

\begin{figure*}[t]
  \centering
  \includegraphics[width=0.95\linewidth]{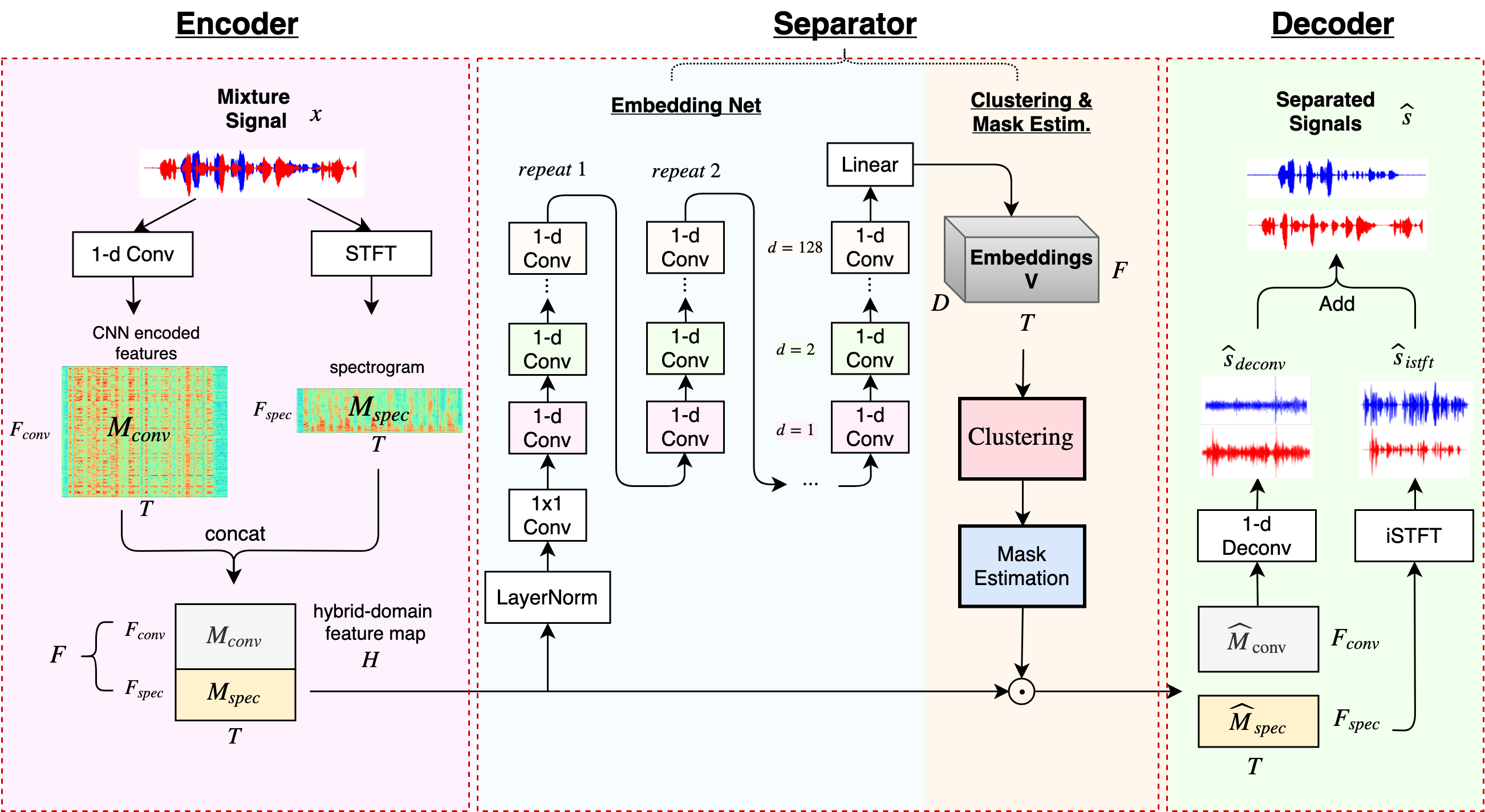}
  \caption{The proposed approach. The encoder extracts features in both time and frequency domains from input signals. The separator includes an embedding network projecting each element of the feature map to a D-dim vector, followed by clustering and mask estimation for each source speaker. The decoder reconstructs the signal waveforms of each source from the masked features.} 
  \label{fig:Architecture}
\end{figure*}

\subsection{Overview of the Proposed Approach}
The overview of the proposed approach is in section 2.1, while the details are in sections 2.2-2.5 .
The proposed approach consists of three processing modules as shown in Figure \ref{fig:Architecture}: an encoder on the left, a separator in the middle, and a decoder on the right. 
\begin{itemize}
    \item The \textbf{Encoder} on the left of Figure \ref{fig:Architecture} encodes the input mixture $x$ into a hybrid-domain 2-dim feature map $H$ with $F=F_{conv}+F_{spec}$ channels and $T$ time frames, where $F_{conv}$ and $F_{spec}$ are respectively the dimensionality of the time and frequency domain features, both of which at each time frame correspond to features extracted from a given small segment of the mixture signal.
    % is composed of two parts: a 1-D convolutional process and short-time Fourier transform. From which we derive a 2-D feature map $M_{conv}$ with $F_{conv}$ channels from the convolutional block and a spectrogram of $F_{spec}$ frequency channels. By concatenating both features along time axis, we can get a hybrid feature map &H& with $(F_{conv}+F_{spec})$ channels, where we use as input to the following networks.
    \begin{equation}
        H = Encode(x)
        % M_{conv} = Conv(x(t)) %Lee: 後面沒用上?
    \end{equation}
    % \begin{equation}
    %     Spec = STFT(x(t)) 
    % \end{equation}
    % \begin{equation}
    %     H = [M_{conv}, Spec]
    % \end{equation}
    \item The \textbf{Separator} in the middle of Figure \ref{fig:Architecture} consists of two parts, an Embedding Network and Clustering plus Mask Estimation. The Embedding Network projects each element in the feature map $H$ onto a D-dimensional space, forming the embeddings $V$. Clustering plus Mask Estimation then follows, from which the masks $M$ for the different speakers are generated, and each element of $H$ is assigned to the speakers based on these masks.
    \begin{equation}
        V = Embed(H)
    \end{equation}
    \begin{equation}
        M = Clust\text{-}Mask(V) %Lee: binary mask? what is the size of  the mask?
    \end{equation}

    \item The \textbf{Decoder} decodes the masked feature map into the estimated time domain signals $\hat{s}$, which is the weighted sum of two signal components respectively obtained from the estimated time and frequency domain features, and $\odot$ denotes element-wise multiplication.
    \begin{equation}
        \hat{s} = Decode( M \odot H)
    \end{equation}
    % where OLA stands for overlap-and-add operation. %Lee: what is this operation?
\end{itemize}

So in this approach we aim to reconstruct time domain signals from both domain features, and directly optimize the signal-to-distortion ratio on the estimated waveform.
%Lee: we do not have definition now?

\subsection{Encoder}

In our cross-domain setting, we utilize both time domain signals and the magnitude spectrogram jointly. The input is the mixture signal $x(t)$ produced by $N$ speakers $s_1(t),...,s_N(t)$, and the corresponding frequency domain representations are obtained by Short-Time Fourier Transform.

\begin{equation}
x(t)=\sum_{i=1}^{N} s_i(t)
\end{equation}
\begin{equation}
X(f,t)=\sum_{i=1}^{N} S_i(f,t)
\end{equation}

As shown in the left part of Figure \ref{fig:Architecture}, the encoder is composed of two parallel procedures: a 1-dim convolutional block and the Short-Time Fourier Transform. To properly integrating the two extracted features from different domains, we set the same window size and the same striding for both domains. This gives a 2-dim feature map $M_{conv}$ with $F_{conv}$ channels from the convolutional block and a spectrogram $M_{spec}$ of $F_{spec}$ frequency channels. We concatenate $M_{conv}$ and $M_{spec}$ along the channel/frequency axis while aligning the time frames, giving a hybrid-domain feature map $H$ with $F=(F_{conv}+F_{spec})$ channels and $T$ time frames. 
% And we combine both feature representation by concatenating CNN encoded features and spectrogram by time indices, where each column (time axis) of the concatenated features are processing on the same time frame. We later refer these concatenated features as \textit{hybrid features}, which consists of $T$ time frames and $F$ feature channels. %Lee: T and F are not in the figure

\subsection{Separator}

The separator has two parts, an embedding network and clustering plus mask estimation.

\subsubsection{Embedding Network}

In order to estimate the speaker assignment for each T-F index on the hybrid-domain feature map $H$, we seek a D-dimensional embedding representing each T-F index. We project the elements in the hybrid-domain feature $H$ to D-dimension embeddings through multiple layers of 1-d dilated convolutional blocks, as shown in the middle block in Figure \ref{fig:Architecture}. Here the "1-d Conv" block in the figure is actually a residual block consisting of a 1x1-conv, a dilated depthwise convolution and a 1x1-conv module, following the prior work \cite{luo2018tasnet}. We stack $B$ residual blocks with the dilation factors of 1,2,...,$2^{B-1}$ and repeat these blocks for $R$ times, %Lee: depthwise?
followed by a linear layer with a D-dimension vector output for each T-F index on the hybrid-domain feature map $H$. This gives the embeddings $V$ for $H$. 

\subsubsection{Clustering and Mask Estimation}

\begin{figure}[t]
  \centering
  \includegraphics[width=\linewidth]{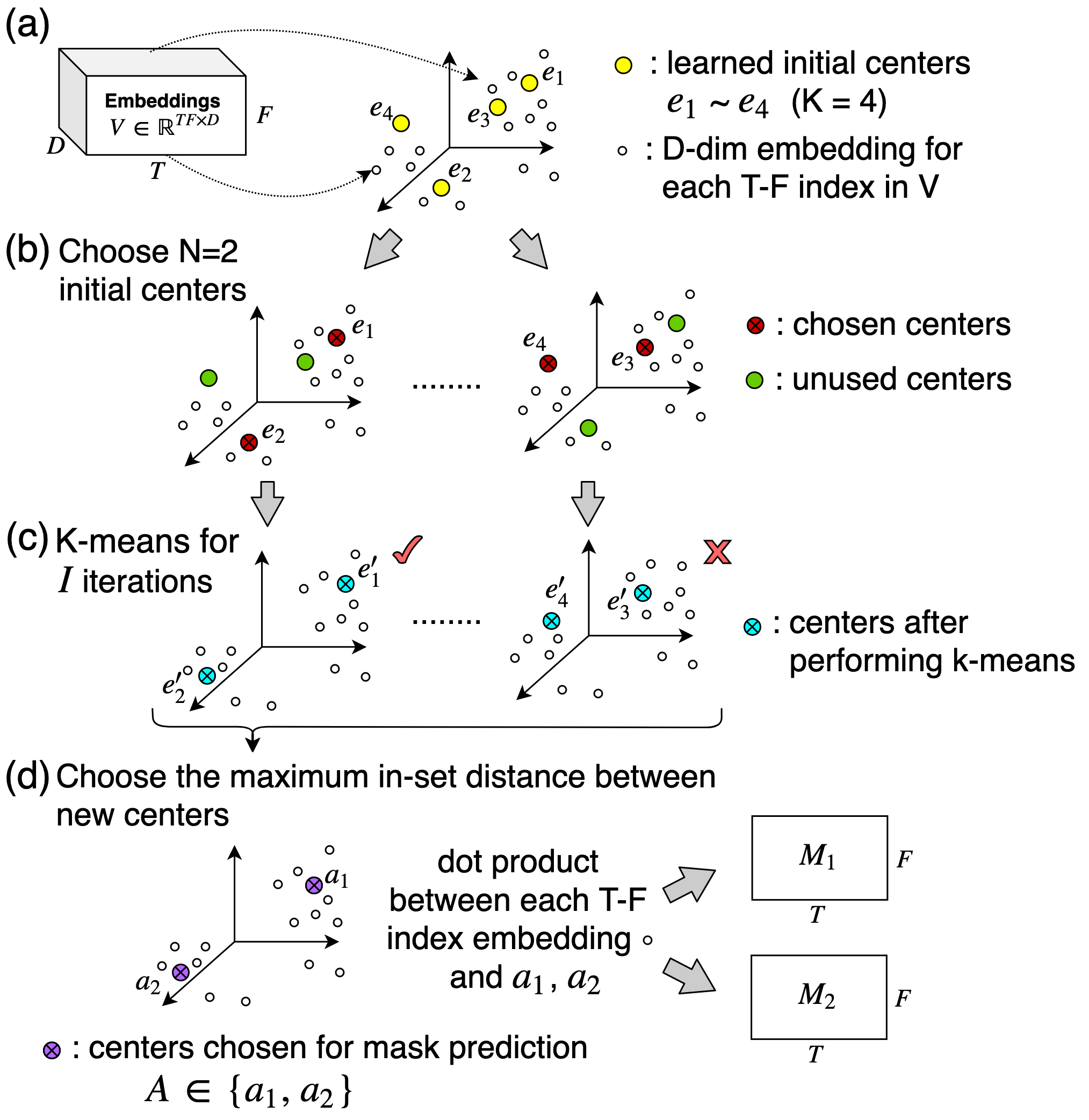}
  \caption{Clustering and mask estimation procedure ( $K=4$ initial centers and $N=2$ speakers shown here).}
  \label{fig:cluster}
\end{figure}

We follow the clustering algorithm previously proposed \cite{luo2018speaker} as shown in Figure \ref{fig:cluster}, starting with $K$ initial centers $e_1, e_2, ..., e_K$ (Figure 2(a)). By arbitrarily choosing $N$ (number of speakers in the mixture) out of the $K$ initial centers (Figure 2(b)), we can get a set of $N$ new centroids by performing k-means clustering for $I$ iterations on the embeddings $V$. Considering all $\binom{K}{N}$ possible selections out of the $K$ initial centers, we can obtain a total of $\binom{K}{N}$ sets of centroids after performing k-means (Figure 2(c)), among which we choose the set of centroids $A$ with the largest in-set distance (Figure 2(d)). The in-set distance is the minimum distance among all pairs of centroids if $N>2$. The masks for each speaker $M \in {\mathbb{R}}^{TF \times N}$ is then estimated by the dot product of the chosen centroids $a_i \in A$ and the embeddings $V$.

\begin{equation}
    M_{i_{t,f}} =V_{t,f} \cdot a_i \qquad \text{for $a_i \in A$}
\end{equation}

% We show the corresponding description of Figure \ref{fig:cluster} in the following:
% \begin{enumerate}
% \item Inputs
%     \begin{enumerate}
%     \item $[e_1,e_2,...,e_k]$: K trainable center points
%     \item $V \in \uppercase{\mathbb{R}}^{TF*D}$ : D-dimensional embeddings
%     \item $N$ : number of speakers
%     \end{enumerate}
% \item K-means Clustering
%     \begin{enumerate}
%     \item Consider all combination where we choose N centers out of K trainable centers, with the total number of combinations being $\binom{K}{N}$. 
%     \item Derive new cluster centroids after performing $I$ iterations of k-means for each combination.
%     \item We choose a set $A$ with the maximum in-set distance among all the combinations with new cluster centroids.
%     \end{enumerate}

% \begin{equation}
%     T_p={A_p}^T{A_p}
% \end{equation}
% \begin{equation}
%     t_p=\max{\left\{ t_{p_{ij}} \right\}}, i \neq j
% \end{equation}
% \begin{equation}
%     A=\mathop{\arg\min}_{A_p}{\left\{ {T_p} \right\}}
% \end{equation}

% \item We derive mask estimation $M \in {\mathbb{R}}^{TF \times N}$ according to the dot product of the chosen centroids $a_i \in A$ and the hybrid embeddings $V$

% \begin{equation}
%     M_{i_{t,f}} =V_{t,f} \cdot a_i \qquad \text{for $a_i \in A$}
% \end{equation}

% \end{enumerate}

\subsection{Decoder}

After multiplying the hybrid-domain feature map $H$ by masks $M$, we disassemble the masked encoded features into their original components: convolutional feature map $\hat{M}_{conv}$ and frequency domain spectrogram $\hat{M}_{spec}$. As shown in Figure \ref{fig:Architecture}, we reconstruct the original signal waveforms from each individual domain: $\hat{s}_{deconv}$ from convolutional feature map and $\hat{s}_{istft}$ from the spectrogram, the former through a deconvolution layer followed by overlap-add method to reconstruct the signals $\hat{s}_{deconv}$; the latter taking the phase of the mixture signal for inverse Fourier Transform to derive $\hat{s}_{istft}$. Weighted sum of the two components with a weight parameter $\alpha$ is then taken as the estimated separated signals $\hat{s}$. 

\begin{equation}
  \hat{s} = \alpha* \hat{s}_{deconv} + (1-\alpha)* \hat{s}_{istft}
  \label{estimate signals}
\end{equation}

\subsection{Training Objective}

We take the negative signal-to-distortion ratio as our training objective.

%Lee: 應該要解釋 \hat{s}_{deconv} \hat{s}_{iSTFT}

\begin{equation}
\begin{aligned}
  \mathcal{L}_{oss} &= - SDR( s,\hat{s}) \\
  &= -10 \log_{10}{\dfrac{\langle s, \hat{s}\rangle^2}{||s||^2 ||\hat{s}||^2-\langle s, \hat{s}\angle^2}}
  \label{time_loss}
\end{aligned}
\end{equation}
where $\langle \cdot,\cdot \rangle$ represents dot product and $||s||^2=\langle s,s \rangle$ denotes the signal power. The training is performed end-to-end, so all components are jointly learned.

\section{Experiments}

\subsection{Datasets}

We evaluated the proposed method on publicly avaliable dataset WSJ0-2mix \cite{hershey2016deep}, which was derived from WSJ0 corpus. The 30hrs of training set and 10hrs of validation set consisted of two-speaker mixtures generated by different speakers from WSJ0 training set  si\_tr\_s mixed at various signal-to-noise ratio between -5 dB and 5 dB. The 5hrs of testing set was similarly generated from WSJ0 validation set si\_dt\_05 and evaluation set si\_et\_05 produced by 16 speakers. All waveforms were resampled to 8 kHz.

We used in addition environmental sounds from Diverse Environments Multichannel Acoustic Noise Database (DEMAND) \cite{thiemann2013diverse} in the test. We resampled all types of background noise from 16kHz to 8kHz, and mixed one arbitrarily chosen type of noise into the mixtures in WSJ0-2mix test set with given signal-to-noise ratio (SNR).

%\subsection{Network architecture setup} %not just network 
\subsection{Experimental Setup}
The window size of the Short-time Fourier Transform (STFT) and the kernel size for the first convolution layer were both 2.5ms, and the square root Hann window was used for STFT. 20-point DFT was performed to extract the 11-dimensional log magnitude feature, combined with the 256-dimensional feature extracted by 1-d conv, and formed 267-dimensional features in the feature map $H$. %這裡感覺怪怪的

For the separator, the feature map $H$ first went through a 1x1 Conv block with 256 filters, followed by 8 residual 1-d Conv blocks, with dilated rate of 1,2,...,128, repeated for 4 times. $D=20$ was chosen to be the embedding dimension following the prior works \cite{hershey2016deep, luo2018speaker, wang2018alternative, wang2018deep} working on speech separation for better comparison. We set $N=4$ initial centers and $I=1$ iteration for k-means following the prior work \cite{luo2018speaker}. %Lee: 引用論文
We also tested the approach without clustering by passing the output of the last 1-d conv in the embedding net through an additional 1x1 convolution block to estimate N masks as done in TasNet \cite{luo2018tasnet}. 
Since the weight $\alpha$ in (8) show minor difference in preliminary experiments, we set $\alpha = 1$ in most of our experiments. 
The networks are trained from scratch on 4-second segments for 100 epochs using Adam algorithm with permutation invariant training \cite{yu2017permutation, kolbaek2017multitalker}. 

%\subsection{Evaluation metrics} %Lee: 這部分不清楚

We evaluate the approach by the signal-to-distortion ratio improvement (SDR$_i$) \cite{vincent2006performance} and the scale-invariant signal-to-noise ratio improvement (SI-SNR$_i$) \cite{luo2018speaker}. 

\subsection{Results}

We report the performance tested on the WSJ0-2mix test set compared to prior works in Table \ref{tab:performance}. We see the approach proposed here (row (g)) offered significant improvement with SDR$_i$ = 16.9 dB and SI-SNR$_i$ = 16.6 dB compared to all prior works including the previous state-of-the-art methods on time domain SDR$_i$ = 15.0 dB (TasNet \cite{luo2018tasnet} in row (e)) and frequency domain SDR$_i$ = 15.6 dB (Sign Prediction Net \cite{wang2018deep} in row (f)). In addition, it is worth mentioning that in contrast to the bidirectional LSTM consisting of 600 units on each direction used in the prior works \cite{wang2018alternative, wang2018deep}, our implementation of depthwise dilated convolution significantly lowered the parameter size by a factor near 1/3.
%Lee: 這裡為何不是跟 Sign predict net 比呢? 它的參數量比 Chimera++ 更多啊???

Table \ref{tab:ablation} shows the ablation studies for the proposed approach. The upper section (I) is for the separator without clustering, so the masks were directly generated from the embeddings. Here in row (1) the Encoder generated only the time domain features (very similar to TasNet \cite{luo2018tasnet} in row (e) of Table \ref{tab:performance}), while in rows (2)(3) the spectrogram was included in addition. From rows (1)(2) we see the spectrogram improved SDR$_i$ by 0.45 dB (from 15.82 to 16.27 dB, first column of rows (1)(2)). Comparing rows (2)(3) we see the value of $\alpha$ didn't make too much difference, or $\alpha=1$ is about good enough. This implied the cross-domain features were very useful in jointly learning the various modules, but the time domain features alone were about adequate to generate the precise waveforms.

The lower Section (II) of Table \ref{tab:ablation} is for the separator including clustering, rows (4)(5) with time domain features only, and rows (6)(7) with cross-domain features. We see adding the clustering module offered improvement in SDR$_i$ of 0.46 dB (from 15.82 to 16.28 dB, rows (4)(5) vs. (1)) for time domain feature alone, and 0.23 to 0.60 dB (from 16.27 to 16.50 or 16.87 dB, rows (6)(7) vs. (2)) for cross-domain features. The spectrogram was certainly useful here too (rows (6)(7) vs. (4)(5)). We also see the choice of the parameter K made the difference for cross-domain features (rows (6)(7)), although this was not clear for time domain features alone (rows (4)(5)). With joint learning including clustering on cross-domain features, we achieve the best result (state-of-the-art to our knowledge) of SDR$_i$ = 16.87 dB in row (7), or 16.9 dB in row (g) of Table \ref{tab:performance}.

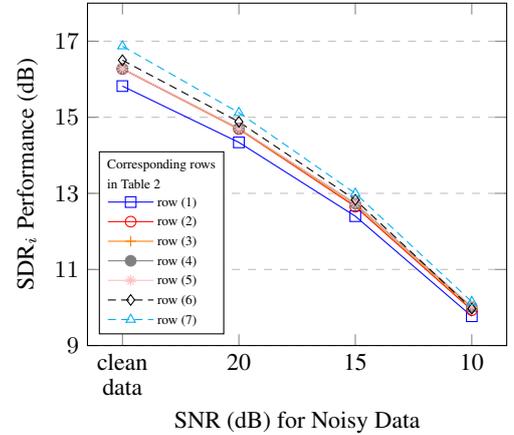
\begin{figure}[H]
\centering
\begin{tikzpicture}
\centering
\begin{axis}[
    xlabel={SNR (dB) for Noisy Data},
    ylabel={SDR$_i$ Performance (dB)},
    ymin=9, ymax=18,
    xtick=data,
    ytick={9,11,13,15,17},
    ylabel near ticks,
    xticklabels={\shortstack{clean \\ data},\text{20},\text{15},\text{10}},
    legend pos=north west,
    ymajorgrids=true,
    grid style=dashed,
    width=0.85\linewidth,
]
\addplot+[color=blue, mark=square, mark size=2pt,]
    coordinates {(0,15.82)(1,14.34)(2,12.40)(3,9.77)};
\label{p1}
\addplot+[color=red, mark=o, mark size=2pt,]
    coordinates {(0,16.27)(1,14.69)(2,12.65)(3,9.93)};
\label{p2}
\addplot+[color=orange, mark=+, mark size=2pt,]
    coordinates {(0,16.28)(1,14.69)(2,12.68)(3,9.96)};
\label{p3}
\addplot+[color=gray, mark=*, mark size=2pt,]
    coordinates {(0,16.28)(1,14.71)(2,12.73)(3,10.01)};
\label{p4}
\addplot+[color=pink, mark=10-pointed star, mark size=2pt,]
    coordinates {(0,16.28)(1,14.71)(2,12.75)(3,10.04)};
\label{p5}
\addplot+[color=black, mark=diamond, mark size=2pt,]
    coordinates {(0,16.50)(1,14.88)(2,12.83)(3,9.97)};
\label{p6}
\addplot+[color=cyan, mark=triangle, mark size=2pt,]
    coordinates {(0,16.87)(1,15.12)(2,13.00)(3,10.15)};
\label{p7}
\end{axis}
\node [draw,fill=white] at (rel axis cs: 0.26,0.3) {
\shortstack[l]{
\tiny Corresponding rows \\
\tiny in Table 2 \\
\tiny \ref{p1} row (1) \\
\tiny \ref{p2} row (2) \\
\tiny \ref{p3} row (3) \\
\tiny \ref{p4} row (4) \\
\tiny \ref{p5} row (5) \\
\tiny \ref{p6} row (6) \\
\tiny \ref{p7} row (7)
}};
\end{tikzpicture}
\caption{Performance degradation with noise level for different configurations in Table 2.} \label{fig:M1}
\end{figure}

The results for noisy mixture test data (clean data training) with SNRs = 20, 15 dB are listed in the middle and the right columns of Table \ref{tab:ablation}. We see all trends observed above remained true, and a degradation of roughly 1.7 dB or less for 20 dB of SNR, and roughly extra 2 dB for extra 5 dB (20-15) of noise. This showed the robustness of the proposed approach. These results are also shown in Figure \ref{fig:M1}, for all rows (1)-(7) of Table \ref{tab:ablation}, plus results for SNR = 10 dB which we were not able to include in Table \ref{tab:ablation}.

\begin{table}[t]
  \caption{$\text{SI-SNR}_i$ and $\text{SDR}_i$ comparison to different prior works tested on WSJ0-2mix dataset. "*" indicates our estimation not written in the original paper.}
  \label{tab:performance}
  \centering
  \begin{tabular}{|c|l|c|c|c|}
    \Xhline{2\arrayrulewidth}
    \multicolumn{2}{|c|}{\textbf{Approaches}} & \textbf{Params} &  \textbf{$\text{SI-SNR}_i$} &  \textbf{$\text{SDR}_i$} \\
    \Xhline{2\arrayrulewidth}
    \multirow{6}{*}{\rotatebox[origin=c]{90}{prior works}}
    & (a) DPCL++ \cite{isik2016single} & 13.6M & 10.8 dB &   -   \\
    & (b) uPIT-ST \cite{kolbaek2017multitalker}& 92.7M &   -  & 10.0 dB \\
    & (c) ADANet \cite{luo2018speaker}&  9.1M & 10.4 dB & 10.8 dB  \\
    & (d) Chimera++ \cite{wang2018alternative}        & 32.9M & 11.5 dB & 12.0 dB \\
    & (e) TasNet \cite{luo2018tasnet}           &  8.8M & 14.6 dB & 15.0 dB \\
    & (f) Sign Pred Net \cite{wang2018deep}  & 36.8M$^*$ & 15.3 dB & 15.6 dB \\
    \hline
    \multicolumn{1}{|c}{} & (g) Proposed &  10M  & 16.6 dB  & 16.9 dB \\
    \Xhline{2\arrayrulewidth}
  \end{tabular}
\end{table}

\begin{table}[t]
  \caption{SDR$_i$ (dB) performance of the proposed approach when the separator included clustering or not (section (\rom{1})(\rom{2})) and the encoder generated time domain features alone or cross-domain features,  with clean or noisy input for different parameters (K: number of cluster centers in Fig 2(a), $\alpha$: weight in (8)).}
  \label{tab:ablation}
  \centering
  \begin{tabular}{|l|c|c||c|c|c|}
   % \hline 
%\toprule[1pt]
    %\shhline
    \Xhline{2\arrayrulewidth}
    \multirow{3}{*}{{\shortstack[c]{Encoder (with \\ Spectr. or not)}}}  & \multirow{3}{*}{K} & \multirow{3}{*}{$\alpha$} &  \multicolumn{3}{c|}{\textbf{SDR$_i$}} \\ \cline{4-6} 
    & & & \multicolumn{1}{c|}{Clean} & \multicolumn{2}{c|}{Noisy Data}  \\ \cline{5-6}
    & & & \multicolumn{1}{c|}{Data} & \multicolumn{1}{c|}{20dB} & \multicolumn{1}{c|}{15dB}  \\
    \Xhline{2\arrayrulewidth}
    \multicolumn{6}{|>{\columncolor[rgb]{0.9,0.9,0.9}}c|}{(\rom{1}) No clustering in Separator} \\
    \hline
    (1) Time        & - & 1   & 15.82 & 14.34 & 12.40 \\ % tas-22
    (2) Time + Freq & - & 1   & 16.27 & 14.69 & 12.65 \\ % hybrid-4 or 24
    (3) Time + Freq & - & 0.5 & 16.28 & 14.69 & 12.68 \\ % hybrid-8 or 28
    % CNN + STFT & - & 0.7 &       &        \\ % hybrid-27
    % CNN + STFT & - & 0.3 &       &        \\ % hybrid-26
    
    \hline
    \multicolumn{6}{|>{\columncolor[rgb]{0.9,0.9,0.9}}c|}{(\rom{2}) With clustering in Separator}    \\
    \hline
    (4) Time        & 2 & 1   & 16.28 & 14.71 &  12.72  \\ % anc-3 
    (5) Time        & 4 & 1   & 16.28 & 14.71 &  12.74  \\ % anc-2 or 6
    % CNN        & 6 & 1   & 16.43 & 14.80 \\ % anc-1 or 5
    (6) Time + Freq & 2 & 1   & 16.50 & 14.88 &  12.83  \\ % hybrid-25
    (7) Time + Freq & 4 & 1   & 16.87 & 15.12 &  13.00  \\ % hybrid-22 
    % CNN + STFT & 6 & 1   & 16.30 & 14.72 \\ % hybrid-21 or 23 or 6
    % CNN + STFT & 6 & 0.5 & 16.29 &   -   \\ % hybrid-15
    %\hline
    \Xhline{2\arrayrulewidth}
  \end{tabular}
\end{table}
%*, x , +, |, o, asterisk, star, 10-pointed star, oplus, oplus*, otimes, otimes*, square, square*, triangle, triangle*, diamond, halfdiamond*, halfsquare*, right*, left*, Mercedes star, Mercedes star flipped, 

\section{Conclusions}

In this paper, we propose to integrate the time and frequency domain features and perform cross-domain joint learning for speech separation. State-of-the-art performance of $SDR_i=16.9$ dB was achieved on the WSJ0-2mix dataset. This verified that the different advantages of the two domains can be well taken, not to mention the correlations between them are useful.

\bibliographystyle{IEEEtran}

\bibliography{mybib}

\end{document}